\documentclass[11pt]{article}
\usepackage{amsmath}
\usepackage{amssymb}
\usepackage{latexsym}
\usepackage{epsfig}

\newcommand{\be}{\begin{equation}}
\newcommand{\ee}{\end{equation}}

\newcommand{\sectiono}[1]{\section{#1}\setcounter{equation}{0}}

\newcommand{\p}{\partial}

\begin{document}

{}~
\hfill\vbox{\hbox{hep-th/0501142}\hbox{MIT-CTP-3586}
}\break

\vskip 3.0cm

\centerline{\Large \bf
Testing Closed String
Field Theory with Marginal Fields}

\vspace*{10.0ex}

\centerline{\large Haitang Yang
and Barton Zwiebach}

\vspace*{7.0ex}

\vspace*{4.0ex}

\centerline{\large \it  Center for Theoretical Physics}

\centerline{\large \it
Massachusetts Institute of Technology}

\centerline{\large \it Cambridge,
MA 02139, USA}
\vspace*{1.0ex}

\centerline{hyanga@mit.edu, zwiebach@lns.mit.edu}

\vspace*{10.0ex}

\centerline{\bf Abstract}
\bigskip
\smallskip

We study the feasibility of level expansion and test the quartic
vertex  of closed string field theory by checking the flatness of
the potential in  marginal directions. The tests, which work out
correctly, require the cancellation of two contributions: one
from an infinite-level computation with the cubic vertex and the
other from a finite-level computation with the quartic vertex. The
numerical results suggest that the  quartic vertex  contributions
are  comparable or smaller than those of level four fields.

\vfill \eject

\baselineskip=16pt

\vspace*{10.0ex}

\tableofcontents

\sectiono{Introduction and summary}

Marginal deformations have provided a useful laboratory to deepen
our understanding of open string field theory. The effective
potential for a marginal field must vanish, but in the level
expansion one sees a potential that becomes progressively flatter
as the level $\ell$  is
increased~\cite{Sen:2000hx,Taylor:2000,Iqbal:2000qg,Coletti:2003ai}.
The marginal operator was taken to be $c \partial X$ and
corresponds to a constant deformation of the $U(1)$ gauge field in
open string theory. The associated spacetime field $a_s$ can be
viewed as a Wilson line parameter. For small $a_s$ the potential
can be expanded in the form
\begin{equation}
g^2 V_\ell(a_s) =\alpha_4 (\ell) \,a_s^4+\mathcal{O}(a_s^6)\,.
\end{equation}
Numerical evidence was found that the coefficient $\alpha_4(\ell)$
decreases as $\ell$ increases. Eventually, $\alpha (\ell) $ was
elegantly shown to be exactly zero as $\ell$ goes to
infinity~\cite{Schnable:2003}. This is, of course, a necessary
condition for the potential to vanish completely at infinite
level. One can also study large marginal deformations and the
relationship between the string field marginal parameter $a_s$ and
the conformal field theory marginal
parameter~\cite{Sen:2004cq,Sen:2000hx}.

In this paper  we use the closed string marginal operator $c\bar
c\p X\bar\p X$ to test closed string field
theory~\cite{Zwiebach:1992ie,Saadi:tb} and to study the
feasibility of level expansion in this theory. In order to do this
we  compute the effective potential for the associated marginal
parameter, which we denote as $a$.  We focus on the leading $a^4$
term in the expansion of this potential for small $a$.  This term
receives two contributions. The first one, $\mathcal{C}(\ell)$,
arises from the cubic vertex by integration of massive fields of
level less than or equal to $\ell$. The second contribution,
$I_4$, arises from the elementary quartic vertex of closed string
field theory and it has no open string field theory analog.
General computations with the quartic vertex are now possible
thanks to the work of Moeller~\cite{Moeller:2004yy}.  If we denote
by $\ell$ the maximum level for the massive closed string states
that are being integrated, the total potential is
\begin{equation}
\label{cslevtheronp} \kappa^2\, \mathbb{V}_{(\ell)}^{tot}(a) =
\bigl( \mathcal{C} (\ell) + I_4~\bigr)\,
 a^4 +\mathcal{O}(a^6)\,.
\end{equation}
It is natural to write the coefficient $\mathcal{C} (\ell)$ as
\begin{equation}
\mathcal{C} (\ell)= \sum_{\ell'=0}^\ell c (\ell') \,,
\end{equation}
where $c(\ell')$ is the contribution from the massive fields of
level $\ell'$. Marginality of $a$ requires that the term in
parenthesis in (\ref{cslevtheronp}) vanishes as $\ell\to \infty$,
or equivalently, that
\begin{equation}
\label{cslevtheron} \mathcal{C}(\infty) + I_4 =
\sum_{\ell'=0}^\infty c (\ell') + I_4~= 0 \,.
\end{equation}
We find strong evidence for  this cancellation by computing $I_4$
and  the coefficients $c (\ell)$ to high level.  This provides a
test of the quartic structure of closed string theory. It is, in
fact, the first computation with the quartic vertex in which there
is a clear expectation that can be checked.  Quartic terms have
been computed earlier, most notably the quartic term in the (bulk)
tachyon potential~\cite{Belopolsky:1994sk,Belopolsky:1994}. In
that case, however, there was no prediction for the magnitude or
the sign of the result. Our present work gives us confidence that
these early computations are correct.

In open string field theory the level of a cubic interaction is
defined to be the sum of the levels of the three states that are
coupled.  It seems likely that the level of cubic closed string
interactions should be defined in the same way.  It is less clear
how to define a level for quartic interactions in such a way that
cubic and quartic contributions may be compared.  Equation
(\ref{cslevtheron}) allows us to do such comparison.  In
particular we can determine the level~$\ell_*$ for which
$c(\ell_*) \sim I_4$.  Since $|c(\ell)|$ decreases with level,
$\ell_*$ is the level at which inclusion of the quartic
interaction seems appropriate. Our results suggest that $\ell_*
\gtrsim 4$.

A puzzle arises in the computations. The value of $\mathcal{C}
(\infty)$ depends only on the cubic vertex of the string field
theory. The value of $I_4$, which must cancel against $\mathcal{C}
(\infty)$, depends on the quartic vertex.  It is well known that
the quartic vertex is not fully determined by the cubic vertex
(although there is a canonical choice). How is it then possible
for the cancellation to work for all four-string vertices
consistent with the cubic vertex? This happens because of two
facts: first, the cubic vertex  determines the boundary of the
region $\mathcal{V}_{0,4}$ of moduli space  that defines the
quartic vertex and, second, the integrand for $I_4$ is a total
derivative and the integral reduces to the boundary of
$\mathcal{V}_{0,4}$.

\medskip
Let's review the organization of this paper. In section 2  we
state our conventions and carry out the computation of the
coefficients $c(\ell)$ for $\ell\leq 4$. In section 3 we obtain a
simple relation between the coefficients $c(\ell)$ and the
analogous coefficients in the open string potential for the
marginal Wilson line parameter. Using this relation and the
results in~\cite{Sen:2000hx,Coletti:2003ai} we  obtain $c(\ell)$
for $\ell\leq 20$.
 With this data we find a
fit for $\mathcal{C} (\ell)$ and extrapolate to find
$\mathcal{C}(\infty)$. This projected value gives an accurate
cancellation against $I_4$, the value of which is calculated in
section 4. In fact, using the unpublished numerical work
of~\cite{moeller,colleti} the cancellation works to five
significant digits. In section 5, we extend our discussion to the
case of the four marginal operators associated with two spacetime
directions. The $O(2)$ rotational symmetry implies the existence
of two independent structures that can enter into the effective
potential to leading (quartic) order in the fields. We compute the
contributions to these structures from the cubic and quartic
string field vertices and again find convincing cancellations. We
offer a discussion of our results  in section 6.

\sectiono{Marginal field potential from cubic interactions}

The bosonic closed string field theory
action~\cite{Zwiebach:1992ie,Saadi:tb} takes the form
\begin{equation}
S=-\frac{2}{\alpha'}\Big(\frac{1}{2} \langle \Psi|c_0^-\,
Q|\Psi\rangle +\frac{\kappa}{3!} \{ \Psi,
\Psi\,,\Psi\} +\frac{\kappa^2}{4!} \{
\Psi,\,\Psi,\Psi,\Psi\} +\cdots\Big).
\end{equation}
Here $Q$ is the BRST operator, $c_0^{\pm}=\frac{1}{2}(c_0\pm\bar
c_0)$, and the string field $|\Psi\rangle$ is a ghost number two
state that satisfies $(L_0-\bar L_0)|\Psi\rangle =0$ and
$(b_0-\bar b_0)|\Psi\rangle =0$. In this paper we only consider
states with vanishing momentum. After setting $\alpha'=2$ and
rescaling $\Psi\to\kappa^{-1}\Psi$, the potential $V= -S$ is given
by
\begin{equation}
\kappa^2 V=\frac{1}{2}\langle \Psi|c_0^-\, Q|\Psi\rangle
+\frac{1}{3!} \{ \Psi,\Psi,\Psi\} +\frac{1}{4!} \{
\Psi,\Psi,\Psi,\Psi\} +\cdots\,\,.
\end{equation}
 We fix the gauge invariance of
the theory using the Siegel gauge $(b_0+\bar b_0)|\Psi\rangle
=0\,.$ The level $\ell$ of a state is defined as $\ell =L_0+\bar
L_0+2\,.$ The  tachyon state $c_1\bar c_1 |0\rangle$ has level
zero and marginal fields have level two. For a convenient
normalization we assume that all spacetime coordinates have been
compactified and the volume of spacetime is equal to one. We then
use $\langle 0| c_{-1}\bar c_{-1} c_0^- c_0^+ c_1 \bar
c_1|0\rangle=1$, or equivalently,
\begin{equation}
\langle  c(z_1) \bar c (\bar w_1)\, c(z_2) \bar c (\bar
w_2)\,c(z_3) \bar c (\bar w_3) \rangle = 2 (z_1-z_2)(\bar w_1-\bar
w_2) (z_1-z_3)(\bar w_1-\bar w_3) (z_2-z_3) (\bar w_2-\bar w_3)\,.
\end{equation}
Since open string field theory uses  $\langle c(z_1)c(z_2)c(z_3)
\rangle_o=$ $ (z_1-z_2)(z_1-z_3) (z_2 - z_3)$ we can write
\begin{equation}
\label{faccorr} \langle  c(z_1)c(z_2)c(z_3) \,   \bar c (\bar
w_1)\bar c (\bar w_2)\bar c (\bar w_3) \rangle = -2\, \langle
c(z_1)c(z_2)c(z_3) \rangle_o \,\cdot\,\langle \bar c (\bar
w_1)\bar c (\bar w_2)\bar c (\bar w_3) \rangle_o\,.
\end{equation}
This closed/open
relation can be used to calculate
the cubic coupling of  three closed string tachyons:
\begin{equation}
\{ c_1\bar c_1,c_1\bar c_1,c_1\bar c_1 \} = 2\cdot \langle
c_1,c_1,c_1\rangle_o\ \cdot \langle \bar c_1,\bar c_1,\bar
c_1\rangle_o = 2\cdot \mathcal{R}^3 \cdot \mathcal{R}^3 = 2
\mathcal{R}^6\,,
\end{equation}
where $\mathcal{R} \equiv {1\over \rho} = {3\sqrt{3}\over 4}
\simeq 1.2990$, $\rho$ is the (common) mapping radius of the disks
that define the three-string vertex, and $\langle
c_1,c_1,c_1\rangle_o = \mathcal{R}^3$ is the coupling of three
tachyons in open string field theory (see, for example,
\cite{Rastelli:2000iu}, eqn.~(5.6)).

\medskip
In this section we only examine quadratic
and cubic interactions. We begin by considering
the effects of the level zero tachyon $t$
on the potential for the (level two) marginal field $a$. The
string field is therefore
\begin{equation}
|\Psi_0\rangle= t\,c_1\bar
c_1|0\rangle + a\,  \alpha_{-1}\bar \alpha_{-1}\, c_1\bar
c_1|0\rangle\, .
\end{equation}
The subscript on the string field indicates the level of the
highest-level {\em massive} field -- in this case zero, because
the tachyon is the only massive state. The kinetic term and cubic
vertex  give the following potential:
\begin{equation}
\label{level 2 potential} \kappa^2
V_{(0)}= -t^2 +
{1\over 3}\, \mathcal{R}^6\, t^3+ \mathcal{R}^2\,  t\,
a^2 =-t^2+\frac{6561}{4096} \, t^3+\frac{27}{16}\,t a^2.
\end{equation}
To find an effective potential for $a$ we fix values of $a$, solve
for the tachyon field, and substitute back in the potential. For
each value of $a$ there are two solutions for the tachyon. One
gives the vacuum branch $V$ while the other one gives the marginal
branch $M$. The tachyon values are
\begin{equation}
t^{V/M}=\frac{8192\pm\sqrt{67108864-544195584 \,a^2}} {39366}\,.
\end{equation}
As  in open string field theory, the marginal parameter is bounded
$|a|\leq 0.3512$. It is not clear how  higher level and higher
order interactions will affect this bound. In the marginal branch
we can expand the potential for small $a$ and find $\kappa^2
V_{(0)}\simeq 0.7119 \,a^4+0.9622\, a^6+\cdots$. The   quartic
coefficient  can be computed directly using the potential in
(\ref{level 2 potential}) without including the $t^3$ term. The
equation for the tachyon becomes linear and we get
\begin{equation}
\label{wegeta}
\kappa^2 V_{(0)} = {3^6\over 2^{10} }a^4 \simeq 0.71191 \,a^4
\quad \to\quad  \mathcal{C} (0) = c(0) = 0.71191\,,
\end{equation}
using the notation described in the introduction. In general, to
find the contribution to $a^4$ from a massive field $M$ we only
need  the kinetic term for $M$ and the coupling $a^2 M$. In terms
of Feynman diagrams we are simply computing a tree graph with four
external $a$'s, two cubic vertices, and an intermediate massive
field.

\medskip

The  string states needed for higher-level computations are built
with oscillators $\alpha_{n\leq -1}, \bar\alpha_{n\leq -1}$ of the
coordinate $X$, Virasoro operators $L'_{m\leq -2}, \bar L'_{m\leq
-2}$ for the remaining coordinates (thus $c=25$), and
ghost/antighost oscillators. We can list such fields
systematically using the generating function:
\begin{eqnarray}
\label{generating function}
f(x,\bar x,y, \bar y)&=&
\prod_{n=1}^\infty \frac{1}{1-\alpha_{-n} x^n}\,
\frac{1}{1-\bar\alpha_{-n} \bar x^n}\,\prod_{m=2}^\infty
\frac{1}{1-L'_{-m} x^m} \,\frac{1}{1-\bar L'_{-m}
\bar x^m}\nonumber\\
&& \cdot\prod_{k=-1\atop k\not=0}^ \infty   (1 + c_{-k} x^k y)
(1+\bar c_{-k} \bar x^k \bar y) \prod_{l=2}^\infty (1+b_{-l} x^l
y^{-1})\, (1+\bar b_{-l} \bar x^l \bar y^{-1})\,.
\end{eqnarray}
A term of the form  $x^{n} \bar x^{\bar n}  y^m \bar y^{\bar m}$
corresponds to a state with $(L_0, \bar L_0) = (n,\bar n)$ and
ghost numbers $(G, \bar G )= (m,\bar m)$.   A massive field $M$ is
relevant to our calculation  if the coupling $M a^2$ does not
vanish. This requires that $M$ have $(G, \bar G) = (1,1)$, an even
number of $\alpha$'s, and an even number of $\bar \alpha$'s.

At level two we get three states: the marginal field itself,
$c_{-1} c_1|0\rangle$, and $\bar c_{-1} \bar c_1|0\rangle$. One
linear combination of the last two is the ghost dilaton and the
other is pure gauge. Since none of the three states couples to
$a^2$, we have $c(2)=0$. At level four $L_0=\bar L_0=1$ and the
coefficients of $(x\bar x y\bar y) $ give all possible  terms.
With the above rule the  set is reduced to
\begin{eqnarray}
|\Psi_4\rangle &=&\phantom{+}f_1\, c_{-1}\bar c_{-1}+\,f_2\,
L'_{-2}\bar L'_{-2} c_1 \bar c_1 +\,(f_3\, L'_{-2} c_1\bar c_{-1}+
\tilde f_3\, \bar L'_{-2} c_{-1}\bar c_1) +\,r_1\,
\alpha^2_{-1}\bar\alpha^2_{-1} c_1\bar c_1
\nonumber\\[0.5ex]
&&+\,(r_2\,\alpha_{-1}^2 c_1\bar c_{-1}+ \tilde r_2\,
\bar\alpha_{-1}^2 c_{-1} \bar c_1)+\,(r_3\,\alpha^2_{-1}\bar
L'_{-2} c_1\bar c_1+ \tilde r_3\,L'_{-2}\bar \alpha^2_{-1}
c_1\bar c_1)\,.
\end{eqnarray}
The corresponding terms in the potential are
\begin{equation}
\begin{split}
\kappa^2 V_{(4)}=& f_1^2 +\frac{121}{432} a^2\,f_1
   +\frac{625}{4}f_2^2 +\frac{15625}{1728} a^2\, f_2
-\frac{25}{2}\big[ f_3^2 +\tilde f_3^2\big] - \frac{1375}{864}
a^2\,\big(f_3+\tilde f_3\big) \\
&+4 r_1^2 + \frac{27}{16} a^2 \,r_1
 -2\big[r_2^2+ \tilde r_2^2\big] + \frac{11}{16}a^2\,
\big(r_2+\tilde r_2\big)  + 25\big[r_3^2+\tilde r_3^2\big] -
\frac{125}{32}a^2\, \big(r_3+\tilde r_3\big)\, ,
\end{split}
\end{equation}
where we used the conservation laws in \cite{Rastelli:2000iu} to
evaluate the cubic interactions. Solving for all the massive
fields and substituting back into $V_{(4)}$ we obtain
\begin{eqnarray}
\kappa^2 V_{(4)} = -\frac{19321}{46656}~ a^4 \simeq
-0.41412\, a^4 \quad \to \quad  c(4) =-0.41412 \,.
\end{eqnarray}
To get the total contribution up to level four we  add the above
to the  result in (\ref{wegeta}):
\begin{eqnarray}
\label{tot48} \kappa^2 \mathbb{V}_{(4)} =\frac{222305}{746496}~
a^4 \simeq 0.29780\, a^4\, \quad \to \quad  \mathcal{C}(4) =
0.29780\,.
\end{eqnarray}
The contribution from level six string fields vanishes because
none of the string fields has even number of $\alpha$'s as well as
even number of $\bar\alpha$'s and satisfies the condition that
$(G,\bar G)=(1,1)$. Therefore $c(6)=0$. We note that
$\mathcal{C}(4)< \mathcal{C}(0)$. To get additional information we
turn to open string field theory.

\sectiono{Contributions to $a^4$ calculated using  OSFT}

As long as we consider closed string states
of ghost number $(1,1)$, work in the Siegel gauge, and
restrict ourselves to quadratic and cubic interactions,
closed string field theory functions as a  kind of product of
two copies of open string field theory. This will enable us
to relate the contributions to the $a^4$ term in the effective
potential to the similar contributions to $a_s^4$ in the
case of open string field theory.

In  classical open string field theory the marginal state
is $|\phi_a\rangle= \alpha_{-1}^Xc_1|0\rangle$ and
the marginal field is called $a_s$~\cite{Sen:2000hx}.
To calculate the quartic potential $a_s^4$ it suffices
to consider
\begin{equation}
\label{osftset} g^2\, V_O (\ell)= \sum_{\ell'=0,2,\dots}^{\ell}
-g^2S_O^{(\ell')}, \hspace{5mm} -g^2\,
S_O^{(\ell)}=\frac{1}{2}\langle
\Phi^{(\ell)}|Q_B|\Phi^{(\ell)}\rangle+ \langle \Phi^{(\ell)},
\phi_a\,, \phi_a\rangle\, a_s^2\,,
\end{equation}
where $O$ is for open string, $g$ is the open string coupling,
$Q_B$ is the open string BRST operator, and $\ell$ is open string
level: $(L_0+1) |\Phi^{(\ell)}\rangle = \ell
|\Phi^{(\ell)}\rangle$. Only even levels contribute because states
of odd level are twist odd and their coupling to $a_s^2$ vanishes.
For each  $\ell$ we  sum over all basis states of ghost number one
in the Siegel gauge:
\begin{equation}
\label{sgopsd} |\Phi^{(\ell)}\rangle\equiv\sum_i
\phi_i^{(\ell)}|\mathcal{O}_i^{(\ell)}\rangle,\hspace{5mm} L_0\,
|\mathcal{O}_i^{(\ell)}\rangle=(\ell-1)\,
|\mathcal{O}_i^{(\ell)}\rangle.
\end{equation}
We will leave out the superscript $\ell$
whenever possible and define
\begin{equation}
m_{ij}\equiv \langle \mathcal{O}_i|c_0|\mathcal{O}_j\rangle,
\hspace{5mm} K_i\equiv \langle \mathcal{O}_i, \phi_a, \phi_a
\rangle\,,
\end{equation}
where $m_{ij}$ is a symmetric nondegenerate matrix. In Siegel
gauge $Q_B=c_0 L_0$ and therefore
\begin{eqnarray}
-g^2\, S_O^{(\ell)}=\frac{\ell-1}{2}\, \,  \phi_i\, m_{ij}\,\,
\phi_j+K_i\, \phi_i\, a_s^2\, ,
\end{eqnarray}
where  summation over repeated $i$ and $j$ indices is implicit.
Using matrix notation, $[M]_{ij} = m_{ij}, [K]_i = K_i, [\phi]_i =
\phi_i$, we readily find the solution for $\phi$ and the value of
the action:
\begin{equation}
\label{OSFT result} \phi = -{1\over \ell-1} (M^{-1} K) \, a_s^2
\quad \to \quad -g^2\, S_O^{(\ell)} = - \,\frac{1}{2(\ell-1)}\,
K^T M^{-1}  K\, a_s^4\,.
\end{equation}
Back in (\ref{osftset}) we have
\begin{equation}
\label{chindef} g^2 V_O(\ell) =\alpha_4(\ell)~ a_s^4= a_s^4
\sum_{\ell'=0,2,\dots }^{\ell} \chi_{\ell'}\,, \quad
\hbox{with}\quad \chi_{\ell'}=-\,\frac{1}{2(\ell'-1)}\, K^T M^{-1}
K\,.
\end{equation}

Let us now turn to closed strings. Because of level matching and
the constraints $(b_0 \pm \bar b_0)\Psi=0$, a closed string field
of level $L_0+ \bar L_0+2 = 2\ell$ in the Siegel gauge can be
written as a sum of factors: $\Psi^{(2\ell)}=\psi_{ij}\,
|\mathcal{O}_i^{(\ell)}\rangle \otimes
|\overline{\mathcal{O}}_j^{(\ell)}\rangle$ where the open string
states are those in (\ref{sgopsd}).  Therefore
\begin{equation}
\langle \mathcal{O}_i \otimes \overline{\mathcal{O}}_j| c_0^-
Q_B|\mathcal{O}_{i'} \otimes \overline{ \mathcal{O}}_{j'} \rangle
= \frac{1}{2} \langle \mathcal{O}_i \otimes
\overline{\mathcal{O}}_j| c_0\bar c_0 (L_0+\bar
L_0)|\mathcal{O}_{i'} \otimes \overline{\mathcal{O}}_{j'} \rangle
= 2(\ell-1)\, m_{ii'}\,m_{jj'},
\end{equation}
where the factor of two in the last step is from the normalization
(\ref{faccorr}). For closed strings the marginal state is
$G=\alpha_{-1}^X\bar \alpha_{-1}^X c_1 \bar c_1|0\rangle$. Since
$G=\phi_a\otimes\bar\phi_a$, the cubic interaction  factorizes:
$\{\mathcal{O}_i \otimes \overline{\mathcal{O}}_j, G,G\} =2\,
K_i\, K_j$. Therefore, up to the order $a^4$, the potential is
calculated from
\begin{equation}
\kappa^2\, \mathbb{V}_{(2\ell)} = \sum_{\ell'=0,2,\dots}^\ell
-\kappa^2S^{(2\ell')}\,,\quad -\kappa^2 S^{(2\ell)}=
\frac{1}{2}\langle \Psi^{(2\ell)}
|c_0^-Q_B|\Psi^{(2\ell)}\rangle+\frac{1}{2}\{ \Psi^{(2\ell)}, G
,G\}\, a^2\,.
\end{equation}
Our earlier comments allow explicit evaluation:
\begin{equation}
-\kappa^2 S^{(2\ell)}= (\ell-1)\, \psi_{ij}\psi_{i'j'}
m_{ii'} m_{jj'}+a^2\, \psi_{ij} K_i
K_j.
\end{equation}
The equation of motion for $\psi_{ij}$ is readily solved:
\begin{equation}
\psi_{ij}=-\frac{1}{2(\ell-1)}\, m^{-1}_{ii'}m^{-1}_{jj'} \,K_{i'}
K_{j'}\, a^2\,.
\end{equation}
Substituting back  into $S^{(2\ell)}$ and using (\ref{chindef}) we
find
\begin{equation}
\label{fdjkoivnj} -\kappa^2 S^{(2\ell)} = -(\ell-1)
\Bigl(-\frac{1}{2(\ell-1)} K^T M^{-1} K\Bigr)^2 \, a^4 \\ =
-(\ell-1)\, \chi_\ell^2 \,a^4\,.
\end{equation}
We  recognize that the contribution to  $a_s^4$ from the open
string fields of level $\ell$ determines  the contribution to
$a^4$ from the closed string fields of level $2\ell$. With the
notation described in the introduction,
\begin{equation}
\label{o8ube}
\kappa^2 \mathbb{V}_{ (2\ell)}
=\mathcal{C}(2\ell) ~ a^4= a^4 \sum_{\ell'=0,2,\dots}^\ell c(2\ell')
\,,\quad \hbox{with} \quad  c(2\ell) =  -(\ell-1) \chi_{\ell}^2 \,.
\end{equation}

The values of $\alpha_4(\ell)$ (recall (\ref{chindef})) for
$\ell=0,2,$ and $4$ can be read from Table~1 of~\cite{Sen:2000hx},
and values up to $\ell=10$ from Table~1 of~\cite{Coletti:2003ai}
(with extra digits provided by~\cite{colleti}).  We reproduce them
in Table~\ref{higher levels}, along with the corresponding values
of $\chi_\ell$. For $\ell=0,2,$ we confirm the closed string
results of section~2. Since fits in powers of $1/\ell$, where
$\ell$ is open string level, accurately describe the behavior of
coefficients in open string effective potentials
\cite{Taylor:2002fy}, we use the data for $\ell=4,6,8,$ and $10$
to fit $\alpha_4$ to $b_0 + b_1/\ell+ b_2/\ell^2$:
\begin{equation}
\label{alpha4pro} \alpha_4(\ell)\simeq -0.00026 +  \frac{0.35681
}{\ell}+ \frac{0.12893 }{\ell^2}\,.
\end{equation}
This is a good fit since $\alpha_4(\ell)$ must vanish for infinite
level.  We now use this fit and  (\ref{o8ube}) to predict the
behavior of $\mathcal{C}(\ell)$ as a function of the closed string
level $\ell$. It follows from (\ref{chindef}) and
(\ref{alpha4pro}) that
\begin{equation}
\chi_{\ell} = \alpha_4(\ell)- \alpha_4(\ell-2)\simeq
-\frac{0.71361}{\ell^2}\,.
\end{equation}
Equation (\ref{o8ube}) then gives
\begin{equation}
\mathcal{C}(2\ell) - \mathcal{C}(2\ell-4)
= -(\ell-1)\chi_\ell^2\simeq  -\frac{0.50925}{\ell^3}\,.
\end{equation}
This equation is consistent with the extrapolation
\begin{equation}
\label{cldffit}
\mathcal{C}(2\ell)\simeq f_0+\frac{0.50925}{(2\ell)^2}\,.
\end{equation}
Comparing with the open string result (\ref{alpha4pro}) we see
that the potential converges faster in closed string theory.
Given (\ref {cldffit})  we  now make a direct fit of $\mathcal{C}$
to $d_0 + d_2/\ell^2 + d_3 /\ell^3$ using the closed string data
in the table for  $\ell=4,6,8,$ and 10:

\begin{equation}
\label{fitac4}
\mathcal{C}(2\ell) \simeq 0.25585 +\frac{0.50581}{(2\ell)^2}
+\frac{1.06366}{(2\ell)^3},
\end{equation}
From this projection we find
\begin{equation}
\label{fpart3}
\mathcal{C}(\infty) \simeq 0.25585\, .
\end{equation}
Recalling (\ref{cslevtheron}), this  number  must be cancelled
by the elementary
quartic contribution  $I_4$.

\begin{table}[ht]
\begin{center}
{\renewcommand\arraystretch{1.1}
\begin{tabular}{|c|c|c|c|c|}
             \hline
            $~~\ell~~$ &
             $\chi_\ell$&$~~\alpha_4(\ell)~ ~$ & $~~c(2\ell)~~$ &
             $\mathcal{C}(2\ell)$\\
             \hline
             0 & $\phantom{-} 0.84375$ & $\phantom{-}0.84375$
             & $\phantom{-}0.71191$ & $0.71191$ \\
             \hline
             2 & $-0.64352$&  $\phantom{-}0.20023$ & $-0.41412$ &
            $0.29780$ \\
             \hline
             4 & $-0.10323$&$\phantom{-} 0.09700$ & $-0.03197$ &
             $0.26583$\\
             \hline
             6& $-0.03420$ &$\phantom{-}0.06280$ &$-0.00585$
             &$0.25998$\\
             \hline
             8 & $-0.01646$ & $\phantom{-}0.04634$ & $-0.00190$ &
             $0.25808$\\
             \hline
             10 & $-0.00962$ & $\phantom{-}0.03672$ & $-0.00083$&
             $0.25725$\\
             \hline
             $\infty$ & -- & $-0.00026$ & -- & $0.25585$\\
             \hline
\end{tabular}}
\end{center}
\caption{\small $\chi_\ell$ and $\alpha_4(\ell)$ give the
contribution of level $\ell$ fields and the total contributions up
to level $\ell$, respectively, to the quartic term in the
potential for the Wilson line parameter $a_s$. The last two
columns give the contribution $c(2\ell) = - (\ell-1) \chi_l^2$  of
closed string fields of level $2\ell$ and the total contributions
$\mathcal{C}(2\ell)$ up to level $2\ell$ to the quartic term in
the potential for the closed string marginal field $a$. The last
row gives the projections from fits.} \label{higher levels}
\end{table}

\sectiono{Elementary contribution to $a^4$}

We now  compute the coupling of four marginal
operators through the four-string elementary vertex of
closed string field theory. If all fields have the
simple ghost structure $\Psi_i=\mathcal{O}_i c_1\bar
c_1|0\rangle$, with $\mathcal{O}_i$ a primary matter operator
of conformal dimension $(h_i, h_i)$,
the elementary  quartic amplitude is~\cite{Belopolsky:1994sk}:
\begin{equation}
\label{quarticrecipe}
\{ \Psi_1, \Psi_2, \Psi_3, \Psi_4\} = -{2\over \pi
}\int_{\mathcal{V}_{0,4}} {dx\, dy~\langle
\mathcal{O}_1(0)\mathcal{O}_2(1)\mathcal{O}_3(\xi)
\mathcal{O}_4(t=0)\rangle\over \rho_1^{2-2h_1}
\rho_2^{2-2h_2} \rho_3^{2-2h_3} \rho_4^{2-2h_4}} \, .
\end{equation}
Here $\rho_i$'s are the mapping radii and the correlator has the
operators inserted at $z=0,1,\xi= x+ i y$, and $t=1/z=0$.  In this
paper all matter operators have dimension $(1,1)$ and the mapping
radii drop out. For the marginal field $a$ the  corresponding
operator is $G= c\bar c \mathcal{O}_{xx}$ with $\mathcal{O}_{xx}
=-\p X\bar\p X$. Using $\langle \partial X (z_1)
\partial X(z_2)\rangle = 1/(z_1-z_2)^{2}$, as well as the
antiholomorphic analog we find
\begin{equation}
\langle \mathcal{O}_{xx}(0)\,\mathcal{O}_{xx}(1)\,
\mathcal{O}_{xx}(\xi)\,\mathcal{O}_{xx}(t=0)\rangle =
\Bigl|1+\frac{1}{\xi^2}+\frac{1}{(1-\xi)^2}\Bigr|^2\,.
\end{equation}
Therefore, the amplitude $\{ G^4\} \equiv \{ G, G,G, G\}$ is
\begin{equation}
\label{introI04}
\{ G^4\}=-\,\frac{2}{\pi}\,I_{0,4} \,,  \quad \hbox{with}
\quad
I_{0,4} \equiv \int_{\mathcal{V}_{0,4}} dx dy\, \Bigl|
1+\frac{1}{\xi^2}+\frac{1}{(1-\xi)^2}\Bigr|^2 \,.
\end{equation}
The moduli $\mathcal{V}_{0,4}$ space is comprised of twelve
regions, a region $\mathcal{A}$ (\cite{Moeller:2004yy},
Fig.\hskip2pt 3) and  eleven regions
 obtained by acting on
$\mathcal{A}$  with the transformations $\xi\to 1-\xi$,
$\xi\to \frac{1}{\xi}$, $\xi\to\bar\xi$, and their
compositions~\cite{Moeller:2004yy}.
The integrand in (\ref{introI04})
is invariant under these transformations, so we
integrate numerically over $\mathcal{A}$ using the quintic fit
provided by Moeller (\cite{Moeller:2004yy}, eqn.\hskip3pt(6.5))
and multiply the result by twelve:
\begin{equation}
I_{0,4} = 12~ \int_{\mathcal{A}} dx dy\, \Bigl|
1+\frac{1}{\xi^2}+\frac{1}{(1-\xi)^2}\Bigr|^2 \simeq 9.65029\,.
\end{equation}
The contribution to the potential from the elementary quartic
interaction is then
\begin{equation}
\label{a4amplit} \kappa^2 V_4=\frac{1}{4!} \{ G^4\}\, a^4= -
{1\over 12\pi} \, I_{0,4}~ a^4 \simeq-\,0.25598\,  a^4\,\quad \to\quad
I_4 = -\,0.25598\,.
\end{equation}
Recalling (\ref{fpart3}), the test indicated in
(\ref{cslevtheron}) gives
\begin{equation}
\mathcal{C}(\infty) + I_4 = 0.25585  - 0.25598  =  -0.00013 \,.
\end{equation}
The cancellation is impressive: the residue is about
$0.05\%$ of $I_4$.

\noindent {\underbar {Best estimates}}: With the most accurate
description of $\mathcal{A}$,
Moeller~\cite{moeller} has calculated the integral $I_{0,4}$ and
his result gives
  $I_4= -0.255\, 872 (\pm 2) $.
Coletti, Sigalov, and Taylor~\cite{colleti} provided us with 
the $\chi_\ell$ for $\ell\leq 150$.
 With this
data we found $\mathcal{C}(300) = 0.255\,876\,575\,2$, a good
estimate of $\mathcal{C}(\infty)$. Fitting $\mathcal{C}$ to
$d_0 + d_2/\ell^2 + d_3 /\ell^3$ 
using
$2\ell=204$ to $2\ell=300$ gives $\mathcal{C}(\infty) = d_0
= 0.255\, 870\, 873\,1$, which agrees with 
$I_4$ to {\em five} significant digits. With the data for $\ell\leq 78$, M. Beccaria
obtained $\mathcal{C}(\infty)=0.255\, 870\, 870\,6\,(3)$ using
Levin acceleration and the BST algorithm~\cite{Beccaria}. 

\medskip
The cancellation confirms that the sign and the normalization in
(\ref{quarticrecipe}) are correct.  This is the same sign that
implies that the quartic tachyon self-coupling is
negative~\cite{Belopolsky:1994,Moeller:2004yy}. We have thus extra
confidence of the correctness of the early calculation of the
quartic term in the tachyon potential.

One can readily see that the integrand in the amplitude $\{G^4\}$
is a total derivative. It is of the form $f(\xi) f(\bar \xi)
d\xi\wedge d\bar \xi$, with
$f(\xi) = 1 + {1\over \xi^2} + {1\over (1-\xi)^2}$.
We then note that $f(\xi) = \partial g(\xi)$ with
$g(\xi) = \xi - {1\over \xi} + {1\over (1-\xi)}$,
well defined in $\mathcal{V}_{0,4}$
since this region excludes $\xi=0, 1,$ and $\xi=\infty$.  Finally,
$f(\xi) f(\bar \xi) d\xi\wedge d\bar \xi = {1\over 2} \, d\bigl(
g(\xi ) f(\bar \xi) d\bar \xi -  f(\xi) g(\bar \xi) d\xi \bigr)$,
which establishes the claim.

\sectiono{A moduli space of marginal deformations}

If multiple marginal operators  define
a moduli space the potential for the corresponding fields
must vanish identically. An instructive example is provided by
the four marginal operators that can be built using the fields
$X$ and $Y$ associated with the
spacetime coordinates $x$ and
$y$. We will study the potential for the string field
\begin{equation}
|\Psi\rangle = \bigl( a_{xx}\alpha_{-1}^X\bar \alpha_{-1}^X +
a_{yy}\alpha_{-1}^Y\bar \alpha_{-1}^Y+ a_{xy}\alpha_{-1}^X\bar
\alpha_{-1}^Y +a_{yx}\alpha_{-1}^Y\bar \alpha_{-1}^X \bigr)\,c_1
\bar c_1|0\rangle\,.
\end{equation}
The marginal fields $a_{xx}$, $a_{yy}$, and
$a_{xy}+a_{yx}$ are metric deformations while
$a_{xy}-a_{yx}$ is a Kalb-Ramond deformation.
The marginal fields are conveniently assembled into
the two-by-two matrix $M$:
\begin{equation}
M=\left(%
\begin{array}{cc}
  a_{xx} & a_{yx} \\
  a_{xy} & a_{yy} \\
\end{array}%
\right).
\end{equation}

It is useful to consider the global $O(2)$ rotational symmetry of
the $(x,y)$ plane. The potential for $M$  should be invariant
under an $O(2)\times O(2)$ symmetry where the first $O(2)$ rotates
the $(\partial X, \partial Y)$ and the second rotates
$(\bar\partial X, \bar\p Y)$. Consider  two rotation matrices $R$
and $S$ ($R^TR =S^TS = 1$). Together they define an element of
$O(2)\times O(2)$ which acts on $M$ as $M\to RMS^T$. To quadratic
order in $M$ there is an invariant $U$ and a {\em quasi}-invariant
$V$:
\begin{equation}
\label{twoinvariants}
U={\rm Tr} (M^T M) \,, \qquad
 V=\det M \,.
\end{equation}
In general $V\to \pm V$, since $R$ and/or $S$ may have determinant
minus one. An example is provided by the parity transformation $S=
\hbox{diag} (1, -1)$. In fact, the $Z_2$ symmetries that arise
because correlators must have even numbers of appearances of
holomorphic and antiholomorphic derivatives of each coordinate are
taken into account by the various parity transformations.  It
follows that to quartic order in the fields we have two
invariants:
\begin{equation}
U^2 \quad \hbox{and}\quad  V^2\,.
\end{equation}
There are no more
independent invariants: the candidate
${\rm Tr}(M^T M M^T M)$ is  equal to  $U^2 - 2 V^2$.

The lowest level potential involves the tachyon and $M$ and
requires no new computation. Since $U$ contains $a_{xx}^2$ the
coefficient coupling $t$ to $U$ is the same as that coupling $t$
to $a^2$ in (\ref{level 2 potential}).  We thus have, as in
(\ref{wegeta}),
\begin{equation}
 \kappa^2
V_{(0)}=\frac{3^6}{2^{10}}\, U^2\simeq 0.7119\, U^2\,.
\end{equation}

At level four  $25$ states enter the computation. We calculated
the effective potential, solved the equations of motion,  and
verified that all terms  assemble into the two anticipated
invariants, giving
\begin{equation}
 \kappa^2 V_{(4)} =
-\frac{19321}{46656}\,U^2+\frac{344}{729} V^2 \simeq -0.4141\,
U^2+0.4719\, V^2.
\end{equation}
The total effective potential up to level four
from the cubic interactions is therefore:
\begin{eqnarray}
\kappa^2 \mathbb{V}_{(4)}=\frac{222305}{746496}\,
U^2+\frac{344}{729} \, V^2 \simeq 0.2978\, U^2+0.4719\, V^2\,.
\label{2dlevel(4,8)}
\end{eqnarray}
At infinite level the  coefficients of  $U^2$ and $V^2$  must be
cancelled by elementary quartic interactions.
 The quartic interactions contribute
\begin{equation}
\label{setup42s}\kappa^2 V_4 =  \gamma_1 \, U^2  + \gamma_2 V^2
= \gamma_1 ( a_{xx}^4 + 2 a_{xx}^2 a_{yy}^2 + \cdots ) + \gamma_2
(a_{xx}^2 a_{yy}^2  + \cdots ) \,.
\end{equation}
where $\gamma_1$ and $\gamma_2$ are constants to be determined.
The value of $\gamma_1$ is determined by our earlier calculation
of the quartic amplitude for $a$. Therefore (\ref{a4amplit}) gives
$\gamma_1 = - I_{0,4} / (12\pi).$ The coefficient of  $a_{xx}^2
a_{yy}^2$ in the potential, to be calculated next,  will give us
the value of $2\gamma_1 + \gamma_2$, from which we
find~$\gamma_2$.

To compute the elementary quartic amplitude  $a_{xx}^2
a_{yy}^2$, we put the operator $\mathcal{O}_{xx}$
 associated
with $a_{xx}$ at $0$ and $1$ and the operator $\mathcal{O}_{yy}$
associated with $a_{yy}$ at $\xi$ and $\infty$. This choice is
arbitrary and does not affect the value of the {\em integrated}
correlator; this is not manifest but is guaranteed by the symmetry
of the four-string vertex and can be checked explicitly. The
matter correlator is:
\begin{equation}
\langle \mathcal{O}_{xx}(0) \mathcal{O}_{xx}(1) \mathcal{O}_{yy}(\xi)
\mathcal{O}_{yy}(t=0)\rangle
=\Big\langle\p X\bar\p X(0)\p X\bar\p X(1) \Big\rangle
\Big\langle\p Y\bar\p Y(\xi)\p Y\bar\p
Y(t=0)\Big\rangle
= 1\,.
\end{equation}
Since the correlator is just one, the amplitude is proportional
to the area $A_{0,4}$ of the region $\mathcal{V}_{0,4}$
viewed as a subset of the $z$ plane (with metric $dzd\bar z$):
\begin{equation}
\{\mathcal{O}_{xx}^2
\mathcal{O}_{yy}^2\}=-\frac{2}{\pi}\,\int_{\mathcal{V}_{0,4}} dx
 dy = -{2\over \pi} ~ A_{0,4}\,.
\end{equation}
Since the contribution of a region $\mathcal{S}$  is the same as
that of $1-\mathcal{S}$, $\overline{\mathcal{S}}$, and
$1-\overline{\mathcal{S}}$ we have
\begin{equation}
\label{areav04}
A_{0,4}=\int_{\mathcal{V}_{0,4}}\hskip-6pt dx dy
= 4\,\Bigl(\, \int_\mathcal{A} +
\int_{1\over \mathcal{A}}
+\int_{\frac{1}{1-\mathcal{A}}}\Bigr) dx dy
= 4\int_\mathcal{A}  dx dy \Bigl( 1 + {1\over |\xi|^4 }
+ {1\over |1-\xi|^4} \Bigr) \simeq 6.0774\,.
\end{equation}
Of course, the integrand for area is a total derivative:
$d\xi\wedge d\bar\xi=\frac{1}{2}d(\xi d\bar \xi-\bar\xi d\xi)$.
Back to the amplitude in question,
\begin{equation}
\kappa^2 V=\frac{6}{4!}\{\mathcal{O}_{xx}^2 \mathcal{O}_{yy}^2\}\,
a_{xx}^2 a_{yy}^2=
-{1\over 2\pi}\, A_{0,4}\, a_{xx}^2 a_{yy}^2\,.
\end{equation}
We thus find:
\begin{equation}
\gamma_2 = -{1\over 2\pi} \Bigl( A_{0,4} - {1\over 3}
I_{0,4}\Bigr)\,.
\end{equation}
Collecting our results, equation (\ref{setup42s}) gives
\begin{equation}
\kappa^2 V_4=-{1\over 12\pi} I_{0,4} \, U^2 - {1\over 2\pi} \Bigl(
A_{0,4} - {1\over 3} I_{0,4}\Bigr) V^2\simeq -0.2560\, U^2
-0.4552\, V^2\,. \label{2d4vertex}
\end{equation}
This quartic contribution cancels
most of the potential in (\ref{2dlevel(4,8)}).
The small residual potential is
\begin{equation}
\kappa^2 V_4^{res} =0.0418\, U^2 +0.0167 \, V^2\,.
\end{equation}
The data is collected in Table~\ref{2D table}. The data for $U^2$
does not represent a new test, higher level computations would
reproduce the result of section 4. The residual coefficient for
$V^2$  is  4\% of the original   contribution. This is evidence
that the infinite-level  computation would give the expected
cancellation.

\begin{table}[ht]
\begin{center}
{\renewcommand\arraystretch{1.1}
\begin{tabular}{|c|c|c|}
            \hline
            \hbox{level} &$U^2$& $V^2$\\
            \hline
            0 & $\phantom{-}0.7119$ &- \\
            \hline
            4 & $-0.4141$ & $\phantom{-}0.4719$ \\
            \hline\hline
            quartic& $-0.2560$&$-0.4552$ \\
            \hline
            residual & $\phantom{-}0.0418$ & $\phantom{-}0.0167$ \\
            \hline
\end{tabular}}
\end{center}
\caption{\small Contributions from the given level to the
coefficients that multiply the invariants $U^2$ and $V^2$ in the
effective potential for the marginal fields. The row ``quartic"
gives the contributions from the elementary quartic interactions.
The last row is the residual quartic potential, obtained after
adding all contributions.} \label{2D table}
\end{table}

\sectiono{Conclusion}

In this paper we have tested the quartic vertex of bosonic closed
string field theory and the concrete description of it provided by
Moeller~\cite{Moeller:2004yy}. The sign, normalization, and region
of integration $\mathcal{V}_{0,4}$ of the quartic interaction were
all confirmed. This region comprises the set of four-punctured
spheres that are not produced by Feynman graphs built with two
cubic vertices and a propagator. Our calculations checked the
flatness of the effective potential for marginal parameters; this
required the cancellation of cubic contributions of all levels
against a finite set of  quartic contributions. We examined this
cancellation in two examples, one with one marginal direction and
one with four marginal directions. In the first one, which we
could carry to high level, the cancellation was very accurate and
became almost perfect once we used additional numerical data
provided by~\cite{moeller,colleti}. In the second example, carried
to low level, the cancellation was less accurate but still
convincing. Amusingly, one of the quartic couplings is equal to
the area of $\mathcal{V}_{0,4}$ in the canonical presentation.

The cancellations were guaranteed to happen if closed string field
theory reproduces a familiar on-shell fact: the S-matrix element
coupling four marginal operators  vanishes. Closed string field
theory breaks this computation into two pieces, one from Feynman
graphs and one from an elementary interaction, thus giving us a
consistency test.  Our test has verified that the moduli space
$\mathcal{M}_{0,4}$ of four punctured spheres is correctly
generated by the Feynman graphs and the region
$\mathcal{V}_{0,4}$.

We found a simple relation between the quartic terms in the closed
string potential for the marginal parameter $a$ and those in the
open string potential for the marginal parameter $a_s$: the
contribution to $a^4$ from closed string fields of level $2\ell$
is given by $c(2\ell) = - (\ell-1)\chi_\ell^2$, where $\chi_\ell$
is the contribution to $a_s^4$ from massive open string fields of
level $\ell$. Since $\chi_\ell\sim1/\ell^2$, we have $c(\ell)\sim
1/\ell^3$.  Convergence is faster in closed string field theory.

We have gleaned some information about level expansion in closed
string field theory by comparing contributions obtained from the
cubic and  quartic vertices. The natural counter here is the level
of the massive fields that are integrated using the cubic vertex
and the propagator. Recalling that the quartic vertex contribution
is $I_4\simeq -0.2560$, the  column for $c$   in Table~\ref{higher
levels} shows that $|c(8)| < |I_4| < |c(4)|$, namely, the quartic
contribution is smaller than that of level four fields and larger
than that of level eight fields.  For the case of the invariant
$V^2$ in Table~\ref{2D table}, the quartic contribution is only
slightly smaller than that from level four fields. These results
indicate that the quartic elementary vertex should be included
once the level of fields reaches or exceeds four. It remains to be
seen if this result holds for other types of computations.

It has been suggested (see~\cite{Okawa:2004rh}, for example) that
quartic interactions may carry an intrinsic level. The level $L_4$
of a quartic coupling could be given by $L_4 =
\alpha+\beta\sum_{i=1}^4 \ell_i$, where $\alpha$ and $\beta$ are
constants to be determined. There is scant evidence for any such
relation, but  we might assume $\beta=1$ and attempt to estimate
$\alpha$ as follows. We learned that $|I_4|$ was bounded by the
contributions from level four and level eight massive fields.
Since the cubic couplings involve one massive field and two
marginal (level two) fields, $I_4$ is bounded by contributions
from level 8 and level 12 interactions.  It would be plausible to
say that $I_4$ carries level 10, in which case $\alpha\sim 2$.
The same logic applied to the computation of the invariant $V^2$
would suggest $\alpha\sim 0$. More work will be necessary to
uncover a reliable formula for the level of the quartic
interaction in closed string field theory.

There are some obvious questions we have not tried to answer. Is
the range of $a$ finite or infinite\,?  The cubic tachyon
contribution suggests the range is finite, but higher level and
higher order interactions could change this result.  There are
also questions related to the zero-momentum dilaton, a physical,
dimension-zero state that fails to satisfy the CFT definition of
marginal state because it is not primary. The dilaton theorem,
however, implies that the dilaton has a flat potential. This
potential is hard to compute because the dilaton is not primary.
This computation, which will appear in a separate
paper~\cite{paper2}, provides new stringent tests of the quartic
string vertex, in particular, of the Strebel quadratic
differential that determines local coordinates at the punctures.
Since the dilaton state  exists for general backgrounds its
potential is part of the universal structure of string field
theory.  The dilaton potential is also an important ingredient in
any complete computation of the potential for the bulk closed
string tachyon.

\smallskip
\noindent {\bf Acknowledgements.} This work is supported in part
by the U.S. DOE 
grant DE-FC02-94ER40818. We are grateful
to M.~Schnabl for discussions and useful suggestions. We also
thank  M.~Beccaria, M.~Headrick, and Y.~Okawa for their comments.
Finally, thanks are due to N.~Moeller as well as E.~Coletti,
I.~Sigalov, and W.~Taylor for making accessible their unpublished
numerical work.

\end{document}